\documentclass[sigconf]{acmart}
\AtBeginDocument{%
  }

\setcopyright{acmlicensed}
\copyrightyear{2026}
\acmYear{2026}
\setcopyright{cc}
\setcctype{by}
\acmConference[ICSE-SEIP '26]{2026 IEEE/ACM 48th International Conference on Software Engineering}{April 12--18, 2026}{Rio de Janeiro, Brazil}
\acmBooktitle{2026 IEEE/ACM 48th International Conference on Software Engineering (ICSE-SEIP '26), April 12--18, 2026, Rio de Janeiro, Brazil}
\acmPrice{}
\acmDOI{10.1145/3786583.3786889}
\acmISBN{979-8-4007-2426-8/2026/04}

\begin{document}

\title{What Does Explainable AI Mean in Practice? Evaluative Requirements from a Longitudinal Clinical Case Study}

\author{Tor Sporsem}
\email{tor.sporsem@ntnu.no}
\orcid{0000-0002-5230-7480}
\affiliation{%
  \institution{NTNU \& SINTEF}
  \city{Trondheim}
  \state{}
  \country{Norway}
}

\author{Stine Rasdal Finserås}
\email{stine.r.finseras@ntnu.no}
\affiliation{%
  \institution{NTNU}
  \city{Trondheim}
  \country{Norway}}

\author{Lars Adde}
\email{lars.adde@ntnu.no}
\orcid{0000-0001-5532-0034}
\affiliation{%
  \institution{St. Olavs Hospital \& NTNU}
  \city{Trondheim}
  \country{Norway}}

\author{Inga Strümke}
\email{inga.strumke@ntnu.no}
\orcid{0000-0003-1820-6544}
\affiliation{%
  \institution{NTNU}
  \city{Trondheim}
  \country{Norway}}

\renewcommand{\shortauthors}{Sporsem \& Finserås \& Strumke}

\begin{abstract}

This paper reports a case study on how explainability requirements were elicited during the development of an AI system for predicting cerebral palsy (CP) risk in infants. Over 18 months, we followed a development team and hospital clinicians as they sought to design explanations that would make the AI system trustworthy. Contrary to the assumption that users need detailed explanations of the inner workings of AI systems, our findings show that clinicians trusted it when it enabled them to evaluate predictions against their own assessments. Our findings show how a simple prediction graph proved effective by supporting clinicians’ existing decision-making practices. Drawing on concepts from both Requirements Engineering and Explainable AI, we use the theoretical lens of Evaluative AI to introduce the notion of \textit{Evaluative Requirements}: system requirements that allow users to scrutinize AI outputs and compare them with their own assessments. Our study demonstrates that such requirements are best discovered through the well known methods of iterative prototyping and observation, making them essential for building trustworthy AI systems in expert domains.
\end{abstract}

\acmArticleType{Review}
\acmCodeLink{https://github.com/borisveytsman/acmart}
\acmDataLink{htps://zenodo.org/link}
\acmContributions{BT and GKMT designed the study; LT, VB, and AP
  conducted the experiments, BR, HC, CP and JS analyzed the results,
  JPK developed analytical predictions, all authors participated in
  writing the manuscript.}


\maketitle

\section{Introduction}
For users to trust AI systems, they often need explanations of the system’s inner workings \cite{barredo_arrieta_explainable_2020}. For example, if an AI system recommends a diagnosis to a doctor, the doctor will likely demand an explanation of how the system arrived at its conclusion before accepting it. The main challenge, however, lies in providing explanations that are understandable to non-technical users \cite{habiba_how_2024}. Addressing this requires first identifying users’ needs for explanations, then translating these needs into system requirements, and finally designing and implementing suitable explanation mechanisms. This process has led to the notion of explainability requirements \cite{chazette_explainability_2020, balasubramaniam_transparency_2023, kohl_explainability_2019}.

The field of explainable Artificial Intelligence (XAI) has largely taken on the responsibility of addressing the challenge of making AI systems trustworthy to users \cite{barredo_arrieta_explainable_2020}. Researchers in this field have developed a wide range of methods to generate explanations for non-interpretable machine learning models, including LIME \cite{lime}, SHAP \cite{shap}, and GradCAM \cite{gradcam}. However, the explanations produced by these techniques often take the form of technical artifacts that are meaningful primarily to developers rather than to end-users \cite{barredo_arrieta_explainable_2020, habiba_how_2024}. When explanations fail to address user needs, trust in the AI system never develops, and users may reject it altogether, undermining the entire development effort \cite{habiba_how_2024}.

Against this backdrop, it is essential to investigate how explainability requirements can be discovered during the development of AI systems. We therefore pose the following research question: \textit{How do developers elicit explainability requirements?} To explore this, we present a longitudinal case study of a development team working on an image recognition machine learning system intended to support clinicians in predicting the risk of cerebral palsy (CP) in infants. Clinicians usually assess the risk of CP themselves, and it is therefore natural that they should require explanations regarding whether and why they should trust an AI system offering an alternative CP risk assessment. By analyzing the developers’ experiences, we provide insights into the strategies and practices used to elicit explainability requirements in a resource-constrained setting. Through this study, we aim to advance the Software Engineering literature on AI system development and to offer practical guidance for practitioners and researchers working in similar contexts.

A main finding of this study is that trustworthiness does not depend solely on explaining the AI system’s inner workings. Instead, we observed that clinicians were willing to trust the system’s predictions once they could \textit{investigate and evaluate} those predictions in detail. We use Tim Miller's conceptual framework of Evaluative AI \cite{miller_explainable_2023} as a theoretical lens to interpret these findings and to explain why the developers in our case were able to establish trust in the AI system’s predictions.


\section{Background}

\subsection{Explainability requirements}
Recent advancements in machine learning have accelerated the integration of AI capabilities into software, significantly transforming software engineering processes \cite{martinez-fernandez_software_2022}. Developing AI software differs from traditional software development in three ways \cite{amershi_software_2019}: First, it is more challenging and complex to discover and manage the data required for AI systems. Second, it demands skills that are often not present in typical software teams. Third, AI components are harder to handle as modular units.

Another significant change is that users now require explanations of how AI systems work, which compels developers to elicit requirements for these explanations \cite{ahmad_requirements_2023}. Studies on explainability requirements first surfaced in the Software Engineering literature in 2019 with Köhl et al.'s \cite{kohl_explainability_2019} attempt to define explainability requirements as: ``A system S must be explainable for target group G in context C with respect to aspect Y of explanandum X." This definition emerged from their observation that there was no systematic and overarching approach to ensure explainability by design, and that it was often not even clear what precisely was meant when demanding explainability from a system. Chazette et al.~\cite{chazette_exploring_2021} criticized the definition of Köhl et al., arguing that it leaves open what exactly is to be explained and that we need a better understanding of what explainability means from a software engineering perspective. Based on a systematic literature review, they propose the following definition: ``A system S is explainable with respect to an aspect X of S relative to an addressee A in context C if and only if there is an entity E (the explainer) who, by giving a corpus of information I (the explanation of X), enables A to understand X of S in C."

In order to make these definitions easier to use, Balasubramaniam et al. \cite{balasubramaniam_transparency_2023} created a model where they are broken down into four components that can be summarized as the following questions : 
\begin{itemize}
    \item To whom to explain? I.e., customers, users, employees, developers, partners, etc.
    \item What to explain? I.e., purpose of the AI system, inputs to the AI system, outputs, data used for training the system, behavior of the system, etc. 
    \item In what kind of situation? I.e., when using the AI system, when building it, when testing it, etc. 
    \item Who explains? I.e., the AI system explains itself, humans explain, an explainability tool explains, etc. 
\end{itemize}

Explainability requirements should be treated as non-functional requirements since they constrain how a system should be implemented rather than specifying a function \cite{chazette_explainability_2020, kohl_explainability_2019, balasubramaniam_transparency_2023, habibullah_non-functional_2021}. However, explanations can be a double-edged sword, as they can have both positive and negative effects depending on design choices made during requirements analysis. This double-edged sword effect was particularly evident in Chazette and Schneider's survey \cite{chazette_explainability_2020}, where participants expressed both strong advantages and disadvantages to receiving explanations. Positive impacts of explanations included better understanding, increased trust, and improved transparency. Negative impacts included reduced usability if poorly designed, information overload, and more complex user interfaces. This highlights the importance of carefully considering how explanations are implemented to avoid unintended negative consequences while achieving the desired benefits. 

There are currently no standard practices for addressing end-user needs for explainability, nor for eliciting explainability requirements \cite{habiba_how_2024}. What the literature does provide, however, is a growing list of challenges in the elicitation process. Chazette and Schneider \cite{chazette_explainability_2020} highlight difficulties such as different users requiring different types of explanations, users wanting control over when explanations are provided, and the need to balance informativeness with the risk of overwhelming users. Balasubramaniam et al.~\cite{balasubramaniam_transparency_2023} expand on these issues, emphasizing that teams must understand the user’s domain and organizational context, and that the purpose of the AI system must be clearly defined before explainability requirements can be meaningfully elicited.

There is a lack of empirical studies on eliciting explainability requirements \cite{ahmad_requirements_2023}, and Balasubramaniam et al.'s \cite{balasubramaniam_candidate_2024} recent literature review, which identified 30 practices for developing explainability requirements, concludes that ``most proposed solutions have not been evaluated in real projects, highlighting the need for empirical studies." Two recent empirical contributions have begun to address this gap: (1) interviews are reported as the most efficient elicitation method when compared to focus groups and surveys, although this study did not investigate ethnographic methods, participatory design, or observational techniques \cite{obaidi_how_2025}; and (2) in the domain of agriculture, age, technology experience, and confidence in using digital systems correlate with the varying need for detail in explanations \cite{girmay_explainability_2025}. Yet these contributions still offer relatively high-level insights. We need in-depth case studies of real development projects to understand how explainability requirements actually emerge and are negotiated in practice, and how elicitation techniques perform under the messy conditions of everyday software engineering. Such case-based evidence can refine, challenge, or qualify existing guidance and is necessary for building practically useful theories of explainability requirements.

\subsection{The Evaluative AI paradigm}
\label{sub:evaluative ai}
In this paper, we use the conceptual framework of \textit{Evaluative AI} as a theoretical lens to interpret our findings and to deepen the understanding of how explainability requirements are elicited. This subsection introduces the concept of Evaluative AI and explains why it provides a suitable perspective for our study.

\begin{table*}[ht]
\caption{Comparison between Explainable AI (XAI) and Evaluative AI}
\label{tab:xai-vs-eai}
\begin{tabular}{p{0.22\linewidth}p{0.35\linewidth}p{0.35\linewidth}}
\toprule
\textbf{Aspect} & \textbf{Explainable AI (XAI)} & \textbf{Evaluative AI} \\
\midrule
Paradigm & \textit{Recommendation-driven}: “recommend and defend” – the system provides an answer and explains why it is correct. & \textit{Hypothesis-driven}: the system provides evidence for and against without giving a single recommendation. \\
\addlinespace
Role of AI & Gives a recommendation and justifies it (contrastive explanations: “Why A instead of B?”). & Provides supporting and refuting evidence for multiple options, leaving the decision to the human. \\
\addlinespace
User Control & Limited; user must react to the AI’s recommendation. & High; user decides which hypotheses/options to explore and when. \\
\addlinespace
Trust and Reliance & Risk of \textit{over-reliance} (blindly following) or \textit{under-reliance} (ignoring AI). & Better calibration of trust, since no single “correct” recommendation is forced. \\
\addlinespace
Cognitive Alignment & Misaligned: assumes people will engage with explanations but often they do not. & Aligned: supports abductive reasoning and Data/Frame theory of sensemaking (natural human decision processes). \\
\addlinespace
Strengths & Increases transparency of AI systems; helpful in simple or large-scale automated decisions. & Enhances expert decision making in medium- to high-stakes, low-frequency decisions where human accountability matters. \\
\addlinespace
Weaknesses & Fosters fixation on one option; explanations often ignored or misused. & Requires more cognitive effort from humans; less suited for fast, low-stakes decisions. \\
\bottomrule
\end{tabular}
\end{table*}

The traditional paradigm in explainable AI (XAI) has been what Miller terms \textit{recommendation-driven decision support} \cite{miller_explainable_2023}: AI systems generate a recommended answer and then attempt to justify it through explanations. The underlying assumption is simple: if the explanation is convincing, users will follow the recommendation; if not, they will reject it. A summary of recent research \cite{miller_explainable_2023}, however, shows that this approach often has little impact on how people actually make decisions. Two problems in particular limit its effectiveness: over-reliance and under-reliance. Over-reliance occurs when users accept the machine’s recommendation even when it is wrong, usually because of unwarranted trust \cite{jacovi_formalizing_2021}. This is linked to automation bias—the belief that “the machine must be right because it is an advanced machine.” Under-reliance is the opposite: users dismiss the machine’s recommendation even when it is correct, typically because of unwarranted distrust \cite{jacovi_formalizing_2021}. This is known as algorithmic aversion—a tendency to reject an output from a machine that one would accept from a human. Both patterns undermine human–AI collaboration and prevent the system from supporting better decision making.

To address these issues, Miller \cite{miller_explainable_2023} introduces the concept of \emph{Evaluative AI}, see Table \ref{tab:xai-vs-eai} for a comparison to XAI. Rather than recommending a single answer, Evaluative AI systems provide evidence for and against a prediction, leaving the final judgment with the human decision maker. This approach explicitly shifts the locus of control: users determine which hypotheses to explore and when, while the machine supplies relevant evidence. It therefore embodies a \textit{machine-in-the-loop} model rather than the more common \textit{human-in-the-loop} framing.

A central motivation for Evaluative AI comes from cognitive science. Miller builds on the Data/Frame theory of sensemaking \cite{klein_dataframe_2007}, which describes how decision makers generate hypotheses, test them, and adjust frames of understanding through abductive reasoning. Evaluative AI is designed to support this natural human reasoning process by allowing exploration of competing hypotheses, weighing evidence for and against, and considering trade-offs. The contrast is the dominating "recommend-and-defend" approach \cite{miller_explainable_2023} which follows a simple logic: the system provides a single recommended answer and then offers reasons to justify why that answer is correct and why alternatives are not. While this can increase transparency, it risks pushing users into either accepting the recommendation without question or rejecting it outright. Evaluative AI avoids providing only one answer and claiming it is the correct one. Instead, it presents evidence both for and against and leaves the decision to the human. Miller argues that this shift reduces fixation on a single outcome, supports expert judgment, and gives users more control over which hypotheses to explore and how to weigh trade-offs \cite{miller_explainable_2023}.

Miller illustrates the paradigm through a diagnostic interface for skin cancer detection. Instead of outputting a single diagnosis, the system filters out unlikely conditions, presents several plausible options (e.g., melanoma, basal cell carcinoma, actinic keratosis), and supplies structured evidence both supporting and contradicting each option. The clinician can then integrate this evidence with their own expertise to reach a decision. This interaction shows how Evaluative AI avoids forcing a recommendation while still guiding attention and supporting judgement.

Importantly, Evaluative AI is not meant to replace explainability but to reframe it: instead of defending machine recommendations, it generates evidence that helps humans critique and refine their own judgements. At the same time, Evaluative AI requires more cognitive effort, and decision makers may prefer the ease of simple recommendations \cite{bucinca_trust_2021}. Yet in medium- and high-stakes contexts where accountability matters, and the decision maker has time to explore options, the added effort can improve decision quality and trust \cite{miller_explainable_2023} Thus, Evaluative AI is proposed as a better fit for expert-driven, high-consequence domains such as medicine, law, and critical infrastructure.

\section{Method}

\subsection{Case description}

This study includes nine infants who took part in a screening program for CP due to complications during pregnancy or childbirth. This screening process can continue until the infants reach the age of six. Notably, nine out of ten infants in the screening program do not have CP, highlighting the potential for sparing parents the nerve-wracking process of screening their child for CP as early as possible. Furthermore, early identification of CP before 6 months age is beneficial for the child's further development due to the plasticity of their young brains, which allows for more effective early interventions and support \cite{novak2017early}.

At present, the screening process for CP is carried out by specially trained clinicians, who are typically able to make a confident diagnosis between 1-2 years of age. The fact that no biomarkers for CP are currently known makes early diagnosis particularly challenging. Clinicians undergo specialized training to recognize CP and accumulate many years of experience to develop this expertise. The ability to predict CP is largely based on intuition and tacit knowledge, as described by one clinician in our study: ``It is a holistic assessment; you get this gut feeling, this suspicion, from observing the child." In Norway, this rare skill is primarily found among clinicians working in the largest hospitals, which often necessitates parents traveling considerable distances to have their child examined. The scarcity of this expertise implies a need for more accessible methods for early CP detection.

Over the course of 17 years, international clinical studies lead by St.\ Olavs University Hospital and and the Norwegian University of Science and Technology (NTNU) has collected 557 videos of infants participating in screening programs. The recent advancements within deep learning have enabled a collaboration between the hospital and the affiliated university to develop an image recognition model using this unique dataset \cite{groos2022development}. The image recognition model predicts CP risk from three-minute videos of infants, recorded either in the clinic or by parents using their smartphones. The model's output is a CP risk prediction ranging from 0 to 1 (representing 0-100\%). 
For a comprehensive description of the model development process and performance scores, refer to~\cite{groos2022development}.

After the CP prediction model was completed, verified, and tested \cite{groos2022development}, researchers began working on explaining the model to uncover the patterns it had identified. However, this endeavor of using XAI methods proved to be exceedingly challenging. Identifying and developing functional XAI methods for the model---as was done continuously in the project, see i.a.~\cite{kimji2025evaluating,kimji2024metrics,tempel2024explaininghumanactivityrecognition,tempel2024chooseexplanationcomparisonshap}---was not sufficient to provide clinicians with satisfactory explanations. After attempting to generate meaningful explanations, the team ultimately decided to develop a software application making the model accessible to clinicians. For the remainder of this paper, we refer to the image recognition model together with the software application as ‘the AI system.’

Now, how would the clinicians respond to this AI system? Would they be willing to actively use it? Would they expect explanations of how it works, and if so, what kind of explanations would meet their needs? This paper reports on a case study of that effort, describing the team’s process, the challenges they encountered, and the approach that ultimately proved effective for eliciting explainability requirements.

\subsection{Case study design}
Case studies form a foundation for creating evidence-based advice for practitioners and contribute to the construction of new theories in academia \cite{stol_teaching_2024}. The objective of this study was to enhance our understanding of the process of eliciting explainability requirements when developing AI systems in an industry setting. To this end, we conducted a longitudinal case study \cite{runeson_guidelines_2008, yin_applications_2011} that followed a development team over 18 months, from the completion of training an image recognition model to the point where a running AI system was deployed and tested by end users.

A longitudinal design strengthens this case study because the elicitation of explainability requirements is not a one-off activity but an evolving process. Case studies benefit from being conducted over time since this can reveal processes, causalities, and changes that would otherwise remain hidden \cite{yin_applications_2011}. In our context, following the development team across several iterations made it possible to trace how explainability requirements emerged, were refined, and became embedded in the system. This longitudinal perspective was essential for understanding not only which requirements were elicited, but also how they were uncovered and transformed into concrete design decisions.

The development team consisted of one software developer, one data scientist enrolled as a software developer, two AI researchers with special competence in XAI, and one product owner (who was building a startup, aiming to commercialize the AI system). The team was primarily collocated, with the software developer occasionally working from a satellite office. Additionally, two additional data scientists, who played key roles in training the image recognition model, served as advisors to the team. A group of three clinicians at the hospital clinic were designated as users and participated in testing the AI system in the clinical setting.

Since all technology used in hospitals in Europe must be approved by the Medical Device Regulatory (MDR), the AI system was not automatically permitted to be tested in clinical settings. Fortunately, the project was funded through the hospital's research budget and the Norwegian Research Council. This gave the work a “research” label, and activities such as experimenting with technology were classified as “research activities.” We had thorough ethical discussions with the head of the clinic on the testing of the AI system and the implications of gathering research data on this testing. In June 2024, the head of clinic accepted that ten infants could be included in testing the AI system (the team ended up recruiting nine). This decision allowed the team to collect a three-minute video of each baby, run the videos through the AI system, and present the CP risk prediction to the clinicians. To minimize the impact on the treatment of the infants, the team was only permitted to present the prediction after a decision for further treatment had been made. Then, for each infant, the clinicians provided feedback in a meeting with the team. During these meetings, the team prioritized what to develop until the next infant arrived for their three-month screening.

\subsection{Data collection and analysis}
To understand the elicitation of explainability requirements, we needed to study both sides of the process: the clinicians who interacted with the AI system in the clinic, and the development team who designed and implemented it. This dual perspective was essential, since requirements emerge not only from user needs but also from how development teams interpret and act on them.

We combined interviews and observations, following the distinction made by Lethbridge et al.\cite{lethbridge_studying_2005} between inquisitive techniques where researchers actively collect information from participants, and observational techniques where researchers passively study how professionals behave in practice. Each method has strengths and drawbacks: interviews can reveal reasoning that is otherwise hidden, but they may be biased or incomplete; observations capture real practices but may introduce effects such as altered behavior when participants know they are being observed. By combining them, we sought to exploit their complementary strengths.

For the development team, we conducted eight interviews and observed ten meetings where requirements and design decisions were discussed. These sessions provided insights into how developers interpreted clinicians’ feedback and translated it into system features. For the clinical side, we carried out ethnographically inspired fieldwork. Since a children’s clinic is a delicate context, only the first author was present in the field. He wore scrubs to blend into the environment and spent entire days with clinicians during infant examinations where the AI system was used. This work amounted to 73 hours of observation, documented in 72 pages of field notes and 35 photos. During these sessions, spontaneous interviews were conducted to capture clinicians’ reflections in situ.

This combination of methods strengthened the study in two ways. First, triangulation across interviews, meeting observations, and ethnographic fieldwork increased the credibility of findings by providing multiple perspectives on the same events \cite{runeson_guidelines_2008}. Second, because explainability needs are often tacit and difficult for experts to verbalize, direct observation of clinicians and developers allowed us to capture how requirements were enacted in practice, not just articulated in interviews. 

We analyzed the data using \textit{narrative analysis}, an approach that focuses on how people make sense of their experiences by structuring them into stories \cite{riessman_doing_1993}. Rather than categorizing data into isolated codes or themes \cite{braun_using_2006}, narrative analysis treats accounts and observations as sequences with beginnings, middles, and ends, where meaning emerges through the unfolding of events, dilemmas, and resolutions. This perspective is particularly useful when studying processes with several actors, because it captures how they connect individual events into a coherent whole \cite{riessman_doing_1993}.

The process of eliciting requirements is well established in the Software Engineering literature. Building on this foundation, we adopted a deductive approach \cite{miles_qualitative_1994} to analyze the elicitation process in our case, interpreting the team’s activities through concepts such as \textit{iterative development} and \textit{incremental delivery}. In contrast, the notion of \textit{explainability requirements} remains under-defined (see Sec. 2.1). To study this emerging concept, we applied an inductive approach, allowing empirical observations to guide how we developed its meaning \cite{miles_qualitative_1994}. To support this dual strategy, we used the Constant Comparison Method \cite{seaman_qualitative_1999}, iteratively comparing insights with data and refining interpretations as the study progressed. We also maintained a research diary to document key events, methodological choices, and reflections, strengthening transparency and reflexivity. Our analysis was guided by an interpretative stance common in empirical SE, which views both software practices and research findings as situated within subjective, socially constructed realities \cite{oates_researching_2022}. The findings in this paper were presented to the clinicians and the development team to rule out any misunderstandings or misrepresentations.
\begin{table}
  \caption{Collected data}
  \label{tab:freq}
  \begin{tabular}{lll}
    \toprule
    Type of data&Amount&Participants\\
    \midrule
    Observations & 10 meetings & Development team\\
    Interviews & 8& Development team\\
    Observations & 73 hours & Clinicians\\
  \bottomrule
\end{tabular}
\end{table}

\section{Findings}

\subsection{\textit{Asking} for explainability requirements}

With a working AI system that predicted CP risk from infant videos, the developers set out to identify explanations that could build trust among clinicians. Without effective explanations, they feared the clinicians would discard the AI system, making years of development effectively useless in practice. Like most startups, the team had limited time and budget, and needed to avoid investing in explanation methods that could turn out to be dead ends. The abundance of methods, combined with the lack of consensus on what constitutes a valid explanation, created further uncertainty.

The team initially tried to ask clinicians directly what kind of explanations they would require to trust the AI system. However, the clinicians struggled to conceptualize AI in an abstract setting. Even with mock-ups, the lack of familiarity with AI hindered them from providing useful responses. Moreover, the busy clinicians often had their minds on real patients and saw the exercise as an interruption. As Ingrid, a data scientist, later summarized:
\begin{quote}
    "It turns out that going to a clinician and asking … ‘What are you looking for?’ … It turns out that clinicians are not able to articulate that."
\end{quote}
Without real insights to guide the development of explanations, the team saw two possible paths forward. One option was to conduct laboratory experiments, developing several candidate explanations and testing them with clinicians. However, this would have been too resource-intensive and was quickly ruled out. The other option was to test the AI system directly in the clinic, observe clinicians’ reactions, and see whether they began requesting explanations. This approach was also challenging: obtaining permission to test on infants was far from trivial (as described in Sec. 3), and the team risked losing clinicians’ trust if they failed to provide adequate explanations. However, the team obtained the acceptance to test the AI system on ten infants by the head of clinic. As Per, the team’s software developer, explained: 
\begin{quote}
    “The questions they start asking can reveal what explanations they need [in order] to trust the model.”
\end{quote}
By analyzing the clinicians’ responses, the team hoped to identify what kinds of explanations were necessary to establish trust in the AI system’s predictions.

\subsection{\textit{Looking} for explainability requirements}
The team created a minimum viable product (MVP), which provided the AI system’s prediction without any accompanying explanations. The rationale behind this approach was that the absence of explanations would provoke clinicians to ask questions, thereby revealing what kinds of explanations they were actually seeking. Yet, infant 1 became a disappointment. The AI system predicted a 1\% risk of CP, which aligned with the clinicians’ conclusion. Rather than sparking discussion, this outcome was met with shrugs and remarks such as “it's good to see that we agree with the machine.”

The situation changed with infant 2. Hanne, a clinician specialized in CP, had just finished her examination. While her overall assessment was optimistic, she and her two colleagues noted some slightly concerning signs.
\begin{quote}
    ``I will call the child back in three months for another examination as a precaution."
\end{quote}
Hanne was then shown the AI system’s prediction: “Total CP risk score: 1\%.” She leaned back with a troubled expression:
\begin{quote}
    ``However, the machine seems very certain, while we are slightly worried. The machine isn't worried at all. This would be an easier pill to swallow if the machine agreed with us."
\end{quote}
She looked out the window, deep in thoughts, before she said:
\begin{quote}
    ``Is it possible for me to see what the model has seen [in the video]? It would be interesting to see if we are thinking alike."
\end{quote}

This moment became the first clear pointer to what explanations the clinicians needed in order to trust the AI system. The team interpreted it as an indication that Hanne did not require technical details of \textit{how} the AI system worked or \textit{what} features it used. Instead, she wanted to compare her own judgment with the AI system’s prediction. This convinced the developers that they had to continue iterating prototypes and observing clinicians’ reactions to uncover explainability requirements. Ingrid, the data scientist reflected: 
\begin{quote}
    “... you need to observe them [the clinicians] as they use the model and see which questions they ask spontaneously.”
\end{quote}

The next step was to figure out how to produce explanations that could be meaningfully compared with the clinicians’ own judgments. To learn more about these judgment practices, the team conducted ethnographic-style observations in the clinic over three days. The idea was that by studying how clinicians arrived at conclusions, they could identify what kinds of explanations would be useful. During these observations, a disagreement arose among clinicians about the CP risk of one infant. One clinician saw signs of CP, while two others did not. This disagreement became a pivotal moment for the team.

\begin{figure}
    \centering
    \includegraphics[width=1\linewidth]{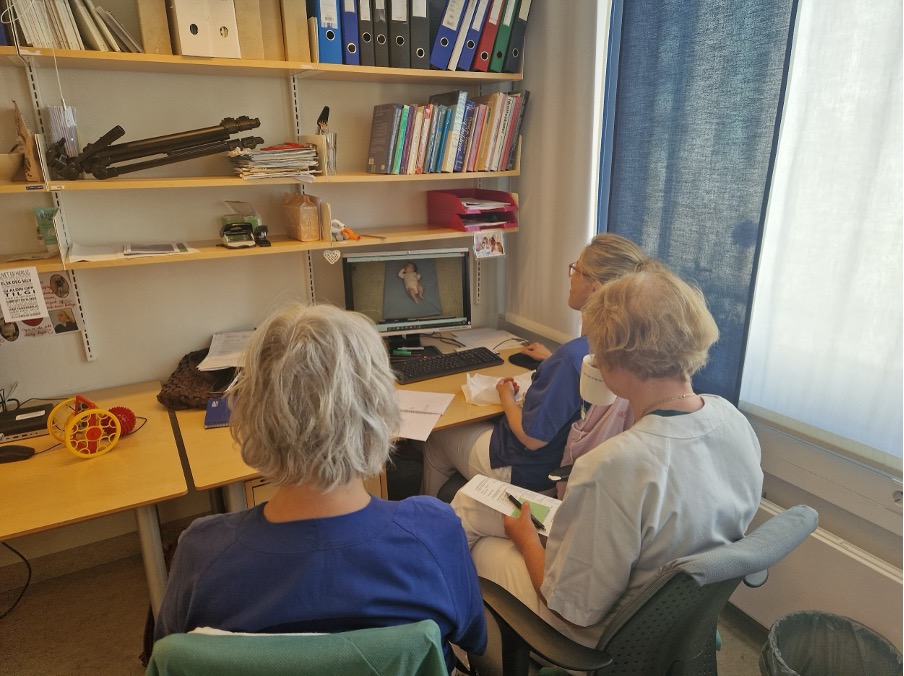}
    \caption{A photograph showing the clinicians watching a video of an infant and disagreeing whether they observe signs of CP or not.}
    \label{fig:gma}
\end{figure}

Figure \ref{fig:gma} shows the three clinicians gathered around the video of the infant in question. The clinician who saw signs of CP moved the time cursor to the section where she noticed them. They watched a 10-second snippet together. Her colleagues looked skeptical, but also uncertain. She replayed the snippet, pointing and making suggestive sounds like “huuum?” Gradually, her colleagues leaned forward, nodded slightly, and began to acknowledge what she was seeing. After a third replay, they nodded in agreement. “I see it,” said one. “I do too,” said the other. They then noted their agreement and decided to schedule a follow-up. 
This scene revealed to the developers how clinicians used tacit knowledge in their common decision-making. Through repeated replays and nonverbal cues, they built consensus without ever explicitly verbalizing what they saw. The developers hypothesized that clinicians might accept a similar process with the AI system: if the system could show which snippets carried higher CP risk, the clinicians could rewatch those sections and judge for themselves whether they agreed.

\begin{figure}
    \centering
    \includegraphics[width=1\linewidth]{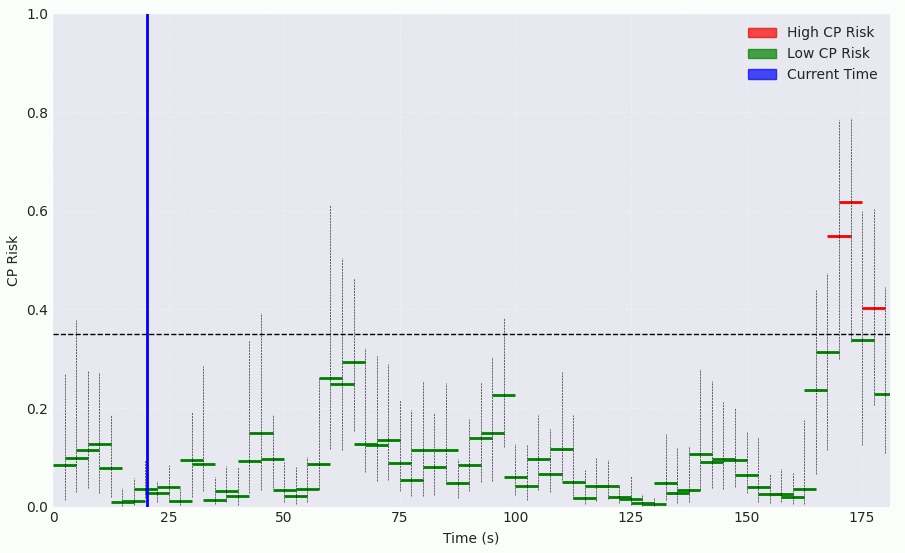}
    \caption{CP risk predictions for each 5-second window of infant 3's video. Most predictions are below the 35\% cut-off, except at the very end.}
    \label{fig:infant3}
\end{figure}

By the time the infant 3 was examined, the developers had produced a graph showing the AI system’s prediction for each 5-second window, see Figure \ref{fig:infant3}. Hanne and two colleagues had just finished their clinical examination and were confident that infant 3 showed no signs of CP. The AI system predicted an 8\% risk. Hanne was satisfied that this was broadly consistent, but she wondered why it did not predict 0 or 1\%. Figure \ref{fig:infant3} illustrates the prediction timeline. Hanne and a colleague immediately investigated the last 30 seconds, where the AI system indicated higher risk. They re-watched the snippet carefully.

\begin{quote}
    ``Look there, yes, the fist is clenched, the foot is positioned like that.''
\end{quote} 
They pointed at the screen, comparing their observations with what the AI system had detected. They described how a clenched fist \textit{could} be a sign of CP, but without other signs, it was usually just a coincidence.
\begin{quote}
    ``Now I can compare what I see with what the model sees."
\end{quote} 
Hanne leaned back, nodded, and remarked “interesting.”

In a later interview, Hanne explained that this ability to judge snippets flagged by the AI system gave her a way to decide whether the prediction was trustworthy in each individual case. As she emphasized, she would ultimately be held accountable for the final decision, and this form of explanation allowed her to evaluate the AI system in relation to her own clinical expertise. The data scientist, Eirik, interpreted Hanne's reaction: 
\begin{quote}
    "Trust can arise through the user’s perceived connection between the model’s output and their own assessments, such as when they were shown the graphs"
\end{quote}
This was initially a surprise to the development team, who had assumed that the clinicians wanted technical explanations about the inner workings of the AI system, provided through XAI-methods. Dag, an XAI specialist on the team explains:
\begin{quote}
    “I was actually surprised that the clinicians didn’t demand detailed explanations. ... What seemed to give them confidence was to see the prediction graph, the outputs over the five-second windows, and then compare that with their own judgment.” 
\end{quote}
The developers later reflected on why they had made this assumption. In interviews, several explained that they themselves felt responsible for whether the AI system produced correct predictions. They wanted to know which patterns the AI system had identified, so that they could decide whether they personally trusted it to make recommendations about infants. Some of them were parents themselves, and Ingrid expressed a genuine wish to make sure the system was safe and reliable.
\begin{quote}
    "And then we have placed the same intentions onto the end-user without actually asking them."
\end{quote}

\subsection{\textit{Validating} explainability requirements}
With infants 4, 5, 6, 7, and 9, the story largely repeated itself. Clinicians consistently included the AI system’s predictions in their judgments by examining the graph with 5-second prediction windows.
\begin{quote}
"I like that it [the AI system] adds to my overall assessment."
\end{quote}
In all these cases, the clinicians generally agreed with the predictions, with only minor differences. Although the development team was pleased that they seemed to have found an effective explanation format, they still wondered what would happen when a real disagreement arose between the AI system’s prediction and the clinicians’ assessment.

\begin{figure}
    \centering
    \includegraphics[width=1\linewidth]{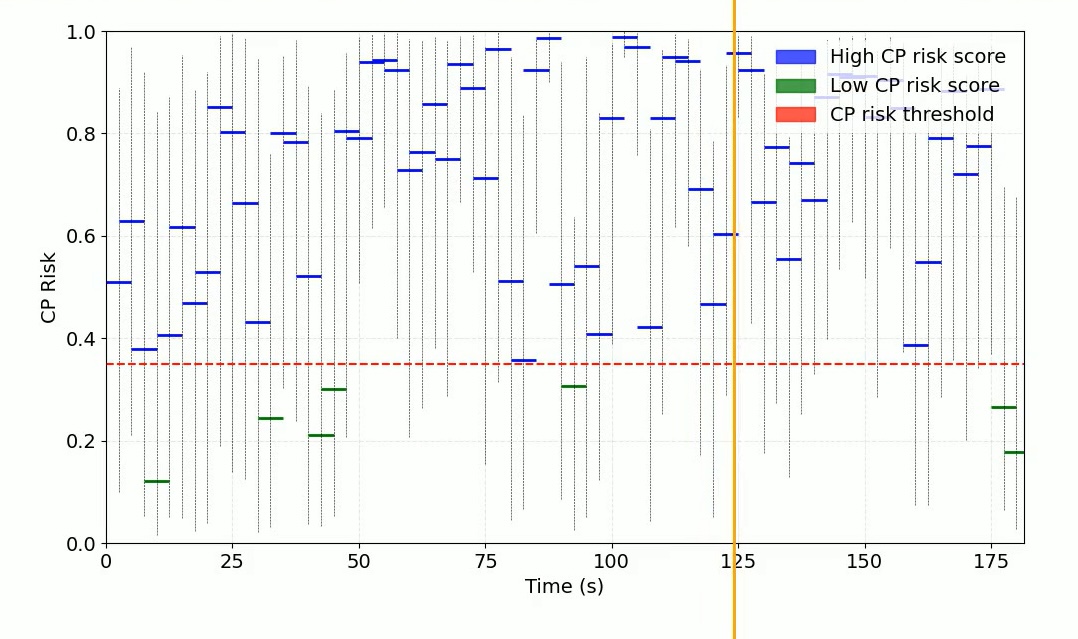}
    \caption{CP risk predictions for each 5-second window for infant 8. Most predictions are above the 35\% cut-off, which disagreed with the clinicians' assessment. The graph differs from Figure \ref{fig:infant3} because it comes from a later iteration.}
    \label{fig:infant8}
\end{figure}

Infant 8 provided exactly such a case: the AI system predicted a 76\% risk of CP, while the clinicians concluded that the risk was low. Hanne studied the graph shown in Figure \ref{fig:infant8}, which prompted her to re-watch the video of infant 8.
\begin{quote}
"My first thought is: what did I miss?"
\end{quote}
The re-watch triggered her carrying out a MOS-analysis (an assessment of the optimality of infant motor repertoires before 5 months age), where infant 8 “did not do very well.” Despite this, the clinicians concluded that they did not see clear signs of CP, but noted that they wanted to keep a closer watch on the child. This demonstrated that the clinicians incorporated the AI system’s prediction into their decision process even when they disagreed with it. The team interpreted this as evidence that they had found a way to activate clinicians’ expertise by enabling them to compare their own judgments with the AI system’s output. In turn, this meant they had succeeded in finding a form of explanation that allowed clinicians to trust the AI system’s predictions. The Software Developer, Per, reflected:
\begin{quote}
"It is not XAI that has convinced our clinicians, it is just the output."
\end{quote}

Despite their initial assumption that explanations from XAI-methods like SHAP, LIME, or GradCAM would be necessary, the team concluded that this enriched output — the 5-second prediction graph — was the most effective way of creating trust and supporting the clinicians’ decision-making. The Data Scientist Eirik: 
\begin{quote}
"In my view, explainability in relation to the final product is about how much clinicians trust the output of the AI, not how transparent the algorithm itself is."
\end{quote}

Our findings demonstrate that the development team was able to identify explainability requirements through iterative experimentation, incremental development, and field testing with observations.
\begin{quote}
    "Each new increment of the application triggered new reflections on what explanations they [the clinicians] needed. Then we tried to meet these needs in the next iteration and tested the effects." 
\end{quote} 
This quote from Per indicates that the familiar software engineering concepts of increments and iterations can efficiently support the discovery and development of explanations for AI systems.

\section{Discussion}
\subsection{Eliciting Explainability Requirements in Practice}

We now return to the research question: How do developers elicit explainability requirements? This case study shows that such requirements were uncovered through observation of users’ reactions. Only through direct observation of the situation in Figure \ref{fig:gma} did the development team gain the necessary understanding of clinicians’ decision-making processes to design the 5-second prediction graph, which proved to be the breakthrough. In contrast, interviews---asking users directly what they needed---did not yield useful results. This outcome stands in clear tension with Obaidi et al.'s \cite{obaidi_how_2025} findings that interviews are the most efficient way of eliciting explainability requirements (a study that explicitly excluded ethnographic and observational techniques), but it aligns with long-standing insights in Software Engineering: for more than 30 years, research has shown that users are often unable to articulate their needs \cite{sporsem_knowns_2023}, and that ethnographic methods are recommended to uncover them \cite{hughes_faltering_1992, hughes_presenting_1995, sommerville_integrating_1993}.

The development team, however, consisted primarily of data scientists with little training or tradition in using qualitative methods for requirements discovery. As the findings illustrate, several were surprised that clinicians were unable to articulate their needs, and they had not anticipated that observation would provide the breakthrough. This reflects critiques of data science as a field, which highlight its limited tradition and lack of systematized methods for qualitative inquiry \cite{tanweer_why_2021}. As data scientists increasingly engage with end users—a trend far more common now than only a few years ago—they also encounter social phenomena, such as trustworthiness in this case. Understanding social phenomena is crucial to ensure that AI systems align with user needs and are not rejected. Addressing these issues requires data scientists to go beyond their dominant quantitative training and engage in \textit{abductive reasoning}, where qualitative and quantitative approaches are combined \cite{meng_discussion_2016}.

Further, users only began to articulate and reflect on their need for explanations \emph{after} interacting with an MVP. Attempts made prior to clinicians' hands-on experience with the AI system yielded little insight into explainability requirements. Providing an early MVP created the conditions for observing clinicians’ interactions and capturing their reflections in real usage situations. Iterations then functioned as requirements probes: each iteration triggered fresh questions and refined the explainability requirements. This approach is well aligned with established Software Engineering practices, where iterative development has long been promoted \cite{boehm_spiral_1986} and is a cornerstone of agile methods \cite{baham_issues_2022}. The strategy of releasing an MVP, experimenting through iterations, and observing user reactions is also widely documented in SE research \cite{lenarduzzi_mvp_2016, duc_minimum_2016} and emphasized in practitioner literature \cite{ries_lean_2011}. As AI becomes increasingly integrated into software systems and data scientists take on roles within software development \textit{organizations} \cite{kim_data_2018, kim_emerging_2016, hukkelberg_exploring_2019, nahar_collaboration_2022}, our findings indicate that these data scientists can use established SE practices to elicit explainability requirements.


\subsection{Introducing Evaluative Requirements}

By applying the conceptual framework of Evaluative AI \cite{miller_explainable_2023} (presented in section \ref{sub:evaluative ai}) to our case study, we argue that the use of the 5-second prediction graphs (Figures \ref{fig:infant3} and \ref{fig:infant8}) is \textit{evaluative} in nature because it enabled clinicians to investigate the AI system’s highlighted predictions separately, and comparing these to their own expert assessments. The graph therefore integrated into their decision-making process: just as they evaluate their colleagues’ assessments, they now evaluated the AI system’s prediction. Rather than being persuaded to trust the AI system's predictions, they used their knowledge to decide for themselves upon inspecting the relevant AI prediction(s). This differs from the dominant model of XAI-assisted decision support, which is built around a recommend-and-defend pattern where disagreement leaves the user with little to critique the predictions or their own ideas \cite{miller_explainable_2023}. Lebovitz et al.~\cite{lebovitz_engage_2022} report a similar finding in diagnostic radiology, where clinicians used "interrogation practices": the AI system produced a prediction, and highlighted the specific regions of CT images flagged as abnormal, enabling clinicians to form their own judgments about its predictions.

We introduce \textbf{Evaluative Requirements}, which define how a system should enable users to scrutinize and compare AI predictions to their own assessments, using their expertise. By grounding Evaluative Requirements in the theoretical lens of Evaluative AI \cite{miller_explainable_2023}, the empirical findings on "interrogation practices" \cite{lebovitz_engage_2022}, and the empirical findings of our case study, we suggest the following definition: \textbf{Evaluative Requirements are system requirements that specify how an AI system must expose and structure evidence, and what interactions it must support, so that users can generate, test, and revise their own hypotheses by comparing AI outputs with their own assessment.} In our case, clinicians did not ask for explanations on \emph{how} the AI system worked; they wanted to see \emph{what} the model “saw” in the video and decide for themselves whether that matched their professional judgment. The 5-second prediction graph is an instance of a feature that satisfies an Evaluative Requirement: it surfaces evidence (segments with predicted high risk) and enables clinicians to re-watch those segments as they desire, allowing them to (re-)evaluate the prediction against their own assessment as long and frequently as they desire.

Discovering Evaluative Requirements, however, is not straightforward, as summarized in Table \ref{tab:eval-req-challenges}. In our study, the developers identified them by observing clinicians in action---watching how they combined expertise, tacit knowledge, and non-verbal communication to reach a conclusion (Figure \ref{fig:gma}). This resonates with Polanyi’s notion of tacit knowledge: expertise is often enacted rather than verbalized, which explains why direct interviews failed to reveal requirements \cite{polanyi_tacit_2009}. Experts may be unable to articulate how their reasoning works, which makes Evaluative Requirements difficult to elicit directly. Instead, the most pragmatic approach appears to be iterative: provide domain experts with prototypes or MVPs, observe how they react, and adapt the system to support their natural decision-making practices.

\begin{table*}[t]
\caption{Challenges in eliciting Evaluative Requirements and their implications for system development practice, as synthesized from our case study.}
\label{tab:eval-req-challenges}
\begin{tabular}{p{0.18\linewidth}p{0.40\linewidth}p{0.35\linewidth}}
\toprule
\textbf{Challenge} & \textbf{Underlying issue} & \textbf{Implications for systems development} \\
\midrule
Context-dependence & Evaluative practices are deeply tied to domain workflows (e.g., clinical examinations, legal reviews). What counts as useful evaluation in one setting may be irrelevant in another, making it risky to reuse requirements across domains. & \textbf{Treat as domain-specific}: run local studies before transferring solutions across fields. \\
\addlinespace
Domain expertise barrier & Understanding evaluative practices requires knowledge of expert reasoning and jargon. Developers without this domain grounding may struggle to recognize which system features genuinely support expert judgement. & \textbf{Emphasize knowledge transfer}: invest in bridging knowledge gaps between domain experts and developers. \\
\addlinespace
Iterative discovery &Evaluative Requirements tend to emerge when users engage with prototypes or MVPs, since these interactions spark reflections and questions. This makes trial-and-error cycles an important part of the process. & \textbf{Prototype iteratively}: embrace experimental cycles and design MVPs to provoke user reactions. \\
\addlinespace
Observation over inquiry & Evaluative needs become visible when experts are observed working with real tasks. However, observing in sensitive environments like hospitals is time-consuming, ethically complex, and requires building trust for access. & \textbf{Observe in context}: use ethnographic or field-inspired methods; plan for ethics approval and stakeholder alignment early. \\
\addlinespace
Developer assumptions & Developers assume that users want technical explanations on the inner workings of AI systems. This leads to building features that users ignore, missing the opportunity to design for evaluation and trust. & \textbf{Validate assumptions}: test through user observation; prioritize evaluative features over technical explanations unless requested. \\
\addlinespace
Efficiency trade-offs & Evaluation requires users to spend extra time (e.g. re-watching videos). While this slows workflows, skipping it risks lack of trust, and system rejection in high-stakes contexts. & 
\textbf{Align with project goals}: if productivity gains are a key success factor, ensure stakeholders understand early that evaluation will slow workflows. \\
\bottomrule
\end{tabular}
\end{table*}

An important consequence of Evaluative AI is that it introduces extra work for users. In our case, clinicians re-watched video segments, compared their own assessments with the AI system’s outputs, and sometimes formulated explanations for why they disagreed. This process is slower than simply accepting or rejecting a recommendation. From a productivity perspective, Evaluative AI can therefore appear to slow down workflows. Yet in high-stakes contexts, this additional time is not wasted---it is precisely what enables trust and clear accountability. At the same time, this tradeoff between speed and decision quality means that Evaluative AI may disappoint stakeholders who expect productivity gains from introducing AI, an expectation that appears to be both widespread and deeply held.

Moreover, Evaluative Requirements may carry a more subtle risk. By giving clinicians the opportunity to examine the AI system's predictions, the system invites them to project their own reasoning patterns onto the system. This can create a sense of alignment that may be non-existing, leading to an illusion of understanding: clinicians may feel assured that they grasp how the system works, even though no genuine transparency into its internal workings is provided. If clinicians believe the system reasons “like they do”, they may give its outputs more weight than is warranted. Instead of treating the system as a fallible tool, they may treat it as a near-peer whose recommendations need only light scrutiny. Conversely, when the system disagrees with them, they may assume they simply “missed something” in their own reasoning, rather than questioning the AI system. 

Lastly, our case shows that Evaluative Requirements are not optional extras that can be added later, but a necessary class of requirements for making AI systems trustworthy in the eyes of users. They strongly influence whether such systems are trusted and adopted, or ultimately abandoned. We argue that Evaluative Requirements are non-functional requirements in the same way that Explainability Requirements are \cite{balasubramaniam_transparency_2023, chazette_explainability_2020, habibullah_non-functional_2021, kohl_explainability_2019}. However, non-functional requirements are often down-prioritized in Agile development \cite{behutiye_management_2020}. It is therefore important that Evaluative Requirements are not treated as secondary concerns in iterative and incremental development, but instead recognized as first-class requirements on par with functional requirements.

\section{Conclusion}

This paper examined how explainability requirements are elicited in the development of AI-enabled software, using a clinical case where an AI system predicts CP risk from infant videos. Our central finding is that trustworthiness in did not depend on technical explanations of AI system internals; instead, clinicians sought the ability to \emph{evaluate} AI system predictions against their own judgement. This motivated the notion of \textit{Evaluative Requirements}: system functions that enable users to interrogate outputs and compare them with their own expert assessments. Recognizing \textit{Evaluative Requirements} as a distinct class of functional requirements helps shift attention from explaining \textit{how} AI systems work to enabling experts to \textit{judge} whether outputs are acceptable in context. This shift is central to making AI systems trustworthy and adoptable to domain experts, such as clinicians.

Methodologically, our study shows that established software engineering practices—iterations, MVPs, and observations—were essential for eliciting both explainability requirements and Evaluative Requirements. This suggests that proven methods from traditional software development can be transferred to AI system development, where trustworthiness is critical. As data scientists enter software organizations, they should be taught these practices, since they seem to provide effective ways to elicit requirements that determine whether AI systems are ultimately trusted and adopted.

\subsection*{Limitations}

\textit{External validity}. This study is based on a single case in one clinical setting with a small number of clinicians and infants, which limits its generalization. Like all case studies, it is subject to the problem of inductive reasoning: even if the findings are convincing in this case, we cannot assume they will hold in other settings. The aim of case studies are therefore not broad generalization, but a deeper understanding of the phenomenon under study. As is common in case study research, the contribution lies in theoretical rather than statistical generalization \cite{runeson_guidelines_2008, yin_applications_2011}. Consequently, the notion of Evaluative Requirements should be validated across other domains.

\textit{Construct validity.} Constructs such as “explainability requirements,” “explainable AI,” and even “AI system” were new to many of the informants in this study, making it challenging to ensure a shared understanding of the terms. In addition, the literature itself lacks consensus on key definitions—for example, what counts as an explanation—making it difficult to capture data consistently and to discuss the concepts scientifically. To help address this issue, two study participants, Strümke and Adde, also took part as co-authors of this article. Concerns about construct validity are common in empirical research, as they relate to whether claims made at the conceptual level are adequately supported by results obtained at the operational level \cite{sjoberg_construct_2023}.

\textit{Researcher/participant dual role.} Two co-authors (Strümke and Adde) also contributed as informants in the study. This dual role carries the risk of expectancy effects, social desirability in discussions, and confirmatory bias during analysis. To mitigate these risks, data collection and analysis were conducted solely by Sporsem and Finserås, while Strümke and Adde contributed during the writing phase. We also grounded our claims in direct quotes and observational evidence. Nevertheless, the possibility of residual bias cannot be ruled out.

\subsection*{Implications for research and practitioners}
We bring the discussion of Evaluative AI \cite{miller_explainable_2023} into the discussions on explainability requirements \cite{kohl_explainability_2019, balasubramaniam_candidate_2024, chazette_explainability_2020}, and bring nuance to what types of explanations developers can look for when developing AI systems. 

As AI become part of more software systems, software developers as a group need to expand their awareness of the new types of requirements that follow. Among these are explainability and Evaluative Requirements. If developers fail to identify such requirements, they risk building systems that, while technically sound, are rejected by their intended users. For this reason, software developers should treat explainability and Evaluative Requirements as part of their responsibility and address them early in the development process. This study also indicates that data scientists can benefit from adopting practices long established in software engineering. However, these qualitative practices are less familiar in the tradition of data science, but they provide a practical path to discovering user needs that are otherwise difficult to articulate.

\subsection*{Future research}

This study opens several avenues for future research on explainability and Evaluative Requirements in AI systems. First, there is a need to better understand what constitutes an MVP when developing AI systems. Does an MVP require a fully trained AI system to generate realistic outputs, or can simplified or mocked-up versions provide sufficient value for eliciting requirements? Systematic studies of different MVP strategies could inform how teams balance cost, speed, and validity in early development stages.  

Second, our case required domain experts with deep knowledge of CP diagnosis to meaningfully evaluate predictions. This raises the question of how Evaluative Requirements differ when the users are novices rather than experts. Future work should investigate how systems can accommodate heterogeneous user groups, ensuring that both novices and experts receive explanations suited to their needs.  

Third, there is a need to formalize and operationalize the concept of \textit{Evaluative Requirements}. At present, the concept remains descriptive and rooted in case observations. Future studies could refine it into frameworks, design guidelines, or measurable criteria that can be systematically applied across domains.  


Finally, replication across domains is essential. The healthcare setting studied here has unique characteristics: high stakes, strong professional accountability, and scarce expertise. Future research should investigate whether the findings transfer to other expert domains such as law, finance, or critical infrastructure, as well as to low-stakes contexts where evaluative effort may be less needed.

\subsubsection*{Acknowledgements}
This research was funded by NTNU, SINTEF, and The Norwegian Research Council (project number: 327146). We sincerely thank the DeepInMotion project team for allowing us to conduct this research. We also thank Torgeir Dingsøyr, Nagadivya Balasubramaniam, and Marjo Kauppinen for their invaluable comments and guidance on early versions of this paper. NTNU and St. Olavs Hospital, may benefit financially from a commercialization of the AI system described in this case through existing intellectual properties; this may include financial benefits to the author Lars Adde of this article.

\bibliographystyle{ACM-Reference-Format}
\bibliography{bib} 


\end{document}